\documentclass[letterp aper,twocolumn,prl,aps,superscriptaddress,amsmath,amssymb,floatfix]{revtex4-2}
\usepackage{mathptmx}
\DeclareMathAlphabet{\mathcal}{OMS}{cmsy}{m}{n}

\usepackage[latin9]{inputenc}
\usepackage{siunitx}
\usepackage{color}
\usepackage{float}
\usepackage{amsmath}
\usepackage{amssymb}
\usepackage{amsthm}
\usepackage{siunitx}
\usepackage[mathscr]{eucal}
\usepackage{graphicx}
\usepackage{esint}
\usepackage[unicode=true,
 bookmarks=true,bookmarksnumbered=false,bookmarksopen=false,
 breaklinks=false,pdfborder={0 0 1},backref=false,colorlinks=true]{hyperref}
\hypersetup{linkcolor=magenta,urlcolor=blue,citecolor=blue,pdfstartview={FitH},hyperfootnotes=false}

\makeatletter

\usepackage{textcomp}
\usepackage{epstopdf}

\pdfpageheight\paperheight
\pdfpagewidth\paperwidth

\@ifundefined{textcolor}{}{%
 \definecolor{BLACK}{gray}{0}
 \definecolor{WHITE}{gray}{1}
 \definecolor{RED}{rgb}{1,0,0}
 \definecolor{GREEN}{rgb}{0,1,0}
 \definecolor{BLUE}{rgb}{0,0,1}
 \definecolor{CYAN}{cmyk}{1,0,0,0}
 \definecolor{MAGENTA}{cmyk}{0,1,0,0}
 \definecolor{YELLOW}{cmyk}{0,0,1,0}
}

\usepackage{xcolor}\usepackage{soul}
\newcommand{\bra}[1]{\ensuremath{\left\langle#1\right|}}
\newcommand{\ket}[1]{\ensuremath{\left|#1\right\rangle}}

\definecolor{blue}{rgb}{0,0,1}
\definecolor{red}{rgb}{1,0,0}
\definecolor{green}{rgb}{0,1,0}

\usepackage{soul}

\makeatother

\begin{document}

\title{Synthetic Gauge Phase in Rydberg Electromagnetically Induced Transparency}

\author{Ya-Dong Hu}
\thanks{These two authors contributed equally to this work.}
\affiliation{Laboratory of Quantum Information, University of Science
and Technology of China, Hefei, Anhui 230026, China}
\affiliation{CAS Center for Excellence in Quantum Information and Quantum Physics,
University of Science and Technology of China, Hefei, Anhui 230026, China}

\author{Yi-Chen Zhang}
\thanks{These two authors contributed equally to this work.}
\affiliation{Laboratory of Quantum Information, University of Science and Technology of China, Hefei, Anhui 230026, China}
\affiliation{CAS Center for Excellence in Quantum Information and Quantum Physics, University of Science and Technology of China, Hefei, Anhui 230026, China}

\author{Qing-Xuan Jie}
\affiliation{Laboratory of Quantum Information, University of Science
and Technology of China, Hefei, Anhui 230026, China}
\affiliation{CAS Center for Excellence in Quantum Information and Quantum Physics,
University of Science and Technology of China, Hefei, Anhui 230026, China}

\author{Hong-Jie Fan}
\affiliation{Laboratory of Quantum Information, University of Science
and Technology of China, Hefei, Anhui 230026, China}
\affiliation{CAS Center for Excellence in Quantum Information and Quantum Physics,
University of Science and Technology of China, Hefei, Anhui 230026, China}

\author{Xiao-Kang Zhong}
\affiliation{Laboratory of Quantum Information, University of Science
and Technology of China, Hefei, Anhui 230026, China}
\affiliation{CAS Center for Excellence in Quantum Information and Quantum Physics,
University of Science and Technology of China, Hefei, Anhui 230026, China}

\author{Dong-Qi Ma}
\affiliation{Laboratory of Quantum Information, University of Science
and Technology of China, Hefei, Anhui 230026, China}
\affiliation{CAS Center for Excellence in Quantum Information and Quantum Physics,
University of Science and Technology of China, Hefei, Anhui 230026, China}

\author{Ya-Nan Lv}
\affiliation{Laboratory of Quantum Information, University of Science
and Technology of China, Hefei, Anhui 230026, China}
\affiliation{CAS Center for Excellence in Quantum Information and Quantum Physics,
University of Science and Technology of China, Hefei, Anhui 230026, China}

\author{Yan-Lei Zhang}
\affiliation{Laboratory of Quantum Information, University of Science
and Technology of China, Hefei, Anhui 230026, China}
\affiliation{CAS Center for Excellence in Quantum Information and Quantum Physics,
University of Science and Technology of China, Hefei, Anhui 230026, China}
\affiliation{Hefei National Laboratory, University of Science and Technology of China,  Hefei 230088, China.}

\author{Xu-Bo Zou}
\affiliation{Laboratory of Quantum Information, University of Science
and Technology of China, Hefei, Anhui 230026, China}
\affiliation{CAS Center for Excellence in Quantum Information and Quantum Physics,
University of Science and Technology of China, Hefei, Anhui 230026, China}
\affiliation{Hefei National Laboratory, University of Science and Technology of China,  Hefei 230088, China.}

\author{Song-Bai Kang}
\affiliation{Innovation Academy for Precision Measurement Science and Technology, Chinese Academy of Sciences, Wuhan 430071, China}

\author{Guang-Can Guo}
\affiliation{Laboratory of Quantum Information, University of Science
and Technology of China, Hefei, Anhui 230026, China}
\affiliation{CAS Center for Excellence in Quantum Information and Quantum Physics,
University of Science and Technology of China, Hefei, Anhui 230026, China}
\affiliation{Hefei National Laboratory, University of Science and Technology of China,  Hefei 230088, China.}

\author{Zhu-Bo Wang}
\email{zbwang@ustc.edu.cn}
\affiliation{Laboratory of Quantum Information, University of Science
and Technology of China, Hefei, Anhui 230026, China}
\affiliation{CAS Center for Excellence in Quantum Information and Quantum Physics,
University of Science and Technology of China, Hefei, Anhui 230026, China}

\author{Chang-Ling Zou}
\email{clzou321@ustc.edu.cn}
\affiliation{Laboratory of Quantum Information, University of Science
and Technology of China, Hefei, Anhui 230026, China}
\affiliation{CAS Center for Excellence in Quantum Information and Quantum Physics,
University of Science and Technology of China, Hefei, Anhui 230026, China}
\affiliation{Hefei National Laboratory, University of Science and Technology of China,  Hefei 230088, China.}

\date{\today}
\begin{abstract}
We demonstrate a synthetic gauge phase in Rydberg electromagnetically induced transparency (EIT) using room-temperature rubidium vapor. By exploiting polarization selection rules in a ladder-type system involving ground, intermediate, and Rydberg states, multiple Zeeman sublevels form closed-loop transitions that acquire a gauge phase. We show that the relative polarization angle between the linearly polarized probe and coupling lasers directly controls this gauge phase, which modulates the EIT transmission and Rydberg state population, consequently controlling the linewidth of EIT due to Rydberg dipole-dipole interactions between atoms. Our approach provides a simple polarization-based method for realizing synthetic gauge physics and manipulating many-body interactions in atomic ensembles without requiring laser cooling and dipole traps.

\end{abstract}
\maketitle

\noindent\textit{1. Introduction.-} Since the 1990s, electromagnetically induced transparency (EIT)~\cite{boller1991observation} in atomic ensembles has attracted great attention, becoming a cornerstone of quantum optics for studying light-matter interactions. EIT is a quantum interference phenomenon in which the absorption of a resonant probe beam is suppressed by the presence of a strong control field that couples to an auxiliary transition. The underlying mechanism involves the destructive interference between different excitation pathways, which creates a narrow transparency window accompanied by a sharp modulation of dispersion. When light-atom interaction is extended from individual atoms to atomic ensembles, EIT gives rise to collective phenomena where the probe photon couples to collective atomic excitations, e.g., spin waves. The non-propagating and long-coherence spin waves enable coherent information storage~\cite{fleischhauer2000darkstate,phillipsStorageLightAtomic2001,wang2019efficient} and drastically reduce the group velocity of light~\cite{kash1999ultraslow}. The universality of EIT as a coherent interference phenomenon has been extended to diverse physical platforms, including optomechanical systems~\cite{safavi-naeini2011electromagnetically}, superconducting quantum circuits~\cite{chu2025slow}, solid-state defects~\cite{turukhin2001observation}, and plasmonic nanostructures~\cite{liu2009plasmonic}.

The textbook treatment of EIT typically employs a simplified three-level model. However, real atoms possess rich internal structure with multiple hyperfine and Zeeman sublevels, offering opportunities for extending EIT physics in two complementary directions. First, incorporating highly excited Rydberg states into the ladder scheme introduces strong, long-range dipole-dipole interactions. This Rydberg EIT platform has revealed fascinating many-body phenomena including dipole blockade~\cite{lukinDipoleBlockadeQuantum2001}, single-photon nonlinearity~\cite{gorshkovPhotonPhotonInteractionsRydberg2011}, and optical bistability~\cite{demeloIntrinsicOpticalBistability2016}, establishing atoms as a powerful medium for quantum simulation~\cite{fontaineObservationBogoliubovDispersion2018,zhangObservationParityTimeSymmetry2016,liuBifurcationTimeCrystals2025,chenContinuousSymmetryBreaking2023,bernienProbingManybodyDynamics2017,ebadiQuantumPhasesMatter2021,schollQuantumSimulation2D2021,blochManybodyPhysicsUltracold2008} and electric field measurements~\cite{zhangMicrowavecoupledOpticalBistability2025,sedlacekMicrowaveElectrometryRydberg2012,wang2025highprecision}. Second, the multilevel structure itself enables multiple coherent pathways connecting the same initial and final states. When the excitation pathways form closed loops, another interference effect emerges from the phase accumulated around the loop, which constitutes a synthetic gauge phase~\cite{an2017direct,chen2024strongly,yuComprehensiveReviewDevelopments2025,huoSolenoidalSyntheticField2014,chen2025interactiondriven,gou2020tunable}, analogous to the Aharonov-Bohm phase acquired by charged particles encircling magnetic flux. Unlike conventional synthetic gauge fields requiring ultracold atoms in optical lattices and tweezer arrays~\cite{gerbier2010gaugea,wu2022manipulating,chengEmergentU1Lattice2024,yuComprehensiveReviewDevelopments2025}, such internal-state loops offer an alternative route to gauge physics in simple atomic ensembles.

In this Letter, we demonstrate a synthetic gauge phase in Rydberg EIT by exploiting both extensions simultaneously, employing Rydberg interactions and closed-loop transitions among Zeeman sublevels. In rubidium vapor at room temperature, we show that the relative polarization angle between the linearly polarized probe and coupling lasers directly controls the gauge phase. As this angle varies, we observe sinusoidal oscillations in both the EIT transmission intensity and linewidth, providing direct evidence of the gauge phase. This mechanism provides an additional control parameter that modulates the Rydberg state population through EIT interference and thus controls the strength of dipole-dipole interactions among Rydberg atoms. The demonstration of gauge phase through simple polarization control provides a knob for exploring new phenomena and applications in atomic ensembles. 

\begin{figure*}
\centering
\includegraphics[scale = 1]{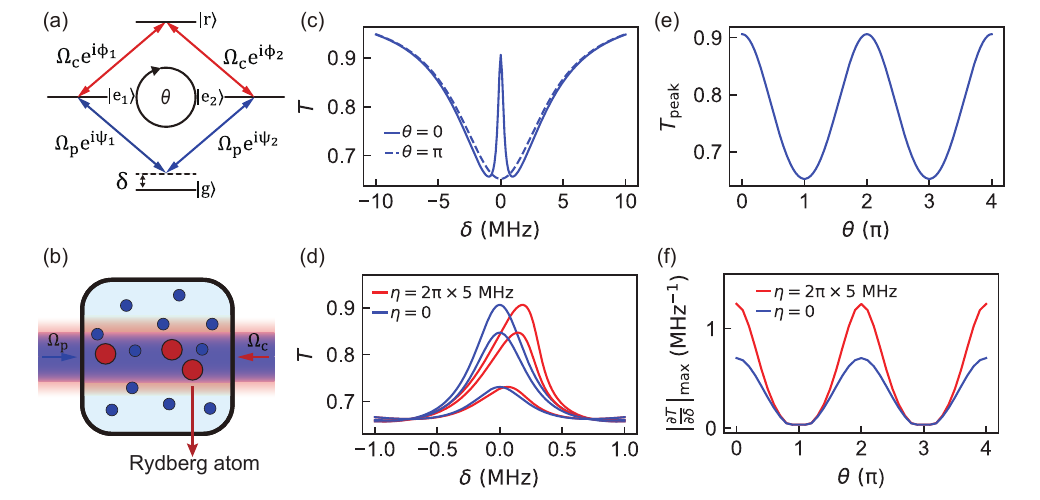}
\caption{(a) Diamond-type energy level configuration for Rydberg EIT, with synthetic gauge phase $\theta$ accumulated in a closed-loop transition. (b) Schematic of counter-propagating probe and control lasers interacting with Rydberg atom ensemble in a vapor cell. (c) Transmission ($T$) spectra for $\theta=0$ (solid line) and $\theta=\mathrm{\pi}$ (dashed line). (d) EIT spectra with (red line) and without (blue line) Rydberg interaction for $\theta=0,\ 0.3\mathrm{\pi,\ 0.6\pi}$ (top to bottom). (e) EIT transmission peak as a function of $\theta$.  (f) Maximum transmission slope versus $\theta$ with (red line) and without (blue line) Rydberg interaction.}
\label{Fig1}
\end{figure*}

\smallskip{}
\noindent\textit{2. Model.-} Figure~\ref{Fig1}(a) illustrates a diamond-type energy level diagram for the Rydberg EIT, where the probe and coupling lasers couple the ground state $\ket{g}$ to a Rydberg state $\ket{r}$ through two intermediate states $\ket{e_1}$ and $\ket{e_2}$ simultaneously. The two ladder-type EIT transition leads to interference, i.e., the absorption of the probe laser and the generated Rydberg state population show destructive or constructive interference depending on the relative phase between the two paths. Equivalently, the diamond-type Rydberg EIT produces a loop of transitions, and  a synthetic magnetic flux emerges in the loop that controls the population transferring among the energy levels. The dynamics of the system can be described by the Hamiltonian 
\begin{align}
\mathcal{{H}} &= {\delta_{\mathrm{eff}}}\ket{g}\bra{g} + \Omega_p\left(e^{i\psi_1}\ket{e_1}\bra{g} + e^{i\psi_2}\ket{e_2}\bra{g}\right) \notag \\
& \quad + \Omega_c\left(e^{i\phi_1}\ket{r}\bra{e_1} + e^{i\phi_2}\ket{r}\bra{e_2}\right) + \text{H.c.}.
\end{align}
Here, $\Omega_p$ and $\Omega_c$ are the Rabi frequencies of probe and coupling lasers, which are set to $2\pi\times$\qty{10}{\kilo\hertz} and $2\pi\times$\qty{100}{\kilo\hertz} in the following simulations. The synthetic gauge phase $\theta$ of this system is defined as the accumulated phase of drives around the loop: $\theta = \phi_2 - \phi_1 + \psi_2 - \psi_1$. We set $\phi_2$, $\psi_1$, and $\psi_2$ to 0 and control the gauge phase by $\phi_1$ in the simulation. The associate decoherence processes are described by the Lindblad terms $\mathcal{L} = \left\{ \kappa_e\ket{g}\bra{e_1}, \kappa_e\ket{g}\bra{e_2}, \frac{\kappa_r}{2}\ket{e_1}\bra{r}, \frac{\kappa_r}{2}\ket{e_2}\bra{r} \right\}$. Here, $\kappa_e$ and $\kappa_r$ denote the decay rates of intermediate states and Rydberg states. We set $\kappa_e = 2\mathrm{\pi}\times$\qty{1.2}{\MHz} and $\kappa_r = 2\mathrm{\pi}\times$\qty{10}{\kilo\hertz} in our simulation.  To study the influence of Rydberg interaction on the synthetic gauge phase, we introduce the mean field theory~\cite{carr2013nonequilibrium} to simplify the many-body dynamics in the atomic ensemble. The model introduces a mean field shift $\delta_{\mathrm{eff}} = \delta + \eta\rho_{rr}$, which contains a term proportional to the Rydberg population $\rho_{rr}$ with a coefficient $\eta$.

\begin{figure*}
\centering
\includegraphics[scale=1]{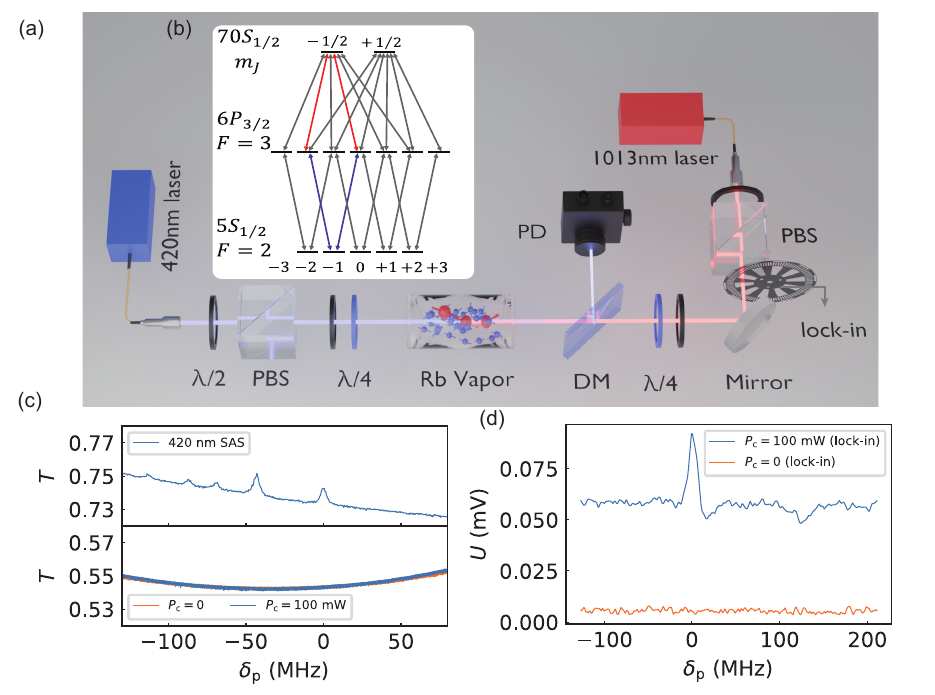}
\caption{(a) Schematic of the experimental setup. PBS: polarization beam splitter; HWP: half-wave plate; QWP: quarter-wave plate; DM: dichroic mirror; PD: photodiode. (b) Relevant energy levels of rubidium atoms. (c) Saturated absorption spectrum (SAS) (top) and the probe absorption profile (bottom) obtained by scanning the probe laser detuning ($\delta_\mathrm{p}$), with and without coupling laser ($P_\mathrm{c}=$\qty{100}{mW}). (d) EIT spectrum versus probe laser detuning, with the coupling laser frequency locked to the $\ket{6P_{3/2}, F=3} \rightarrow \ket{70S_{1/2}}$ transition. EIT signals  are detected using lock-in amplification with modulation acting on the coupling laser while the frequency of the coupling laser and the intensity of the probe laser are fixed. The probe and coupling laser powers are \qty{1}{mW} and \qty{100}{mW} (\qty{0}{mW}), with horizontal and left circular polarizations, respectively. The EIT signal is acquired as the voltage $U$.}
\label{Fig2}
\end{figure*}

We numerically calculated the transmission of probe based on the steady-state solution of the master equation (Fig.~\ref{Fig1}(b)), with the effective atomic density derived from the saturation vapor pressure of $^{87}\text{Rb}$ at \qty{370}{K}. When $\theta=0$ and $\eta=0$, the system exhibits a standard structure of ladder-type EIT, exhibiting a narrow transparent window within a broad absorption valley, as illustrated in Fig.~\ref{Fig1}(c). When $\theta=\pi$, the transparent window disappears. As we tune the gauge phase more finely, the results in Fig.~\ref{Fig1}(d) show a gradual change of the transmission peak value. In Fig.~\ref{Fig1}(e), we extract the transmission peak $T_\mathrm{peak}$ of EIT window, showing sinusoidal oscillations with $\theta$. Furthermore, Fig.~\ref{Fig1}(d) compares the result of the system with and without Rydberg interaction. We find that when we have a finite Rydberg interaction strength (here we set $\eta=2\mathrm{\pi}\times$\qty{5}{MHz}), the Hamiltonian becomes nonlinear, leading to asymmetric transmission lineshapes and a shift of the resonance frequency. We characterize the nonlinear effect by calculating the maximum slope of the transmission $\left| {\mathrm{\partial}T}/{\mathrm{\partial}\delta} \right|_{\mathrm{max}}$. Figure~\ref{Fig1}(f) shows that when we have a finite interaction strength, the maximum slope oscillates with a larger amplitude as a function of the gauge phase $\theta$, indicating the interplay between gauge phase and atom-atom interactions. 

\smallskip{}
\noindent\textit{3. Experimental setup.-} 
Figure~\ref{Fig2}(a) shows the experimental setup for studying the Rydberg EIT in $^{87}$Rb atomic vapor (\qty{370}{K}). The setup comprises two laser sources: a weak \qty{420}{nm} probe laser coupleing the ground state $\ket{g}=\ket{5S_{1/2}, F=2}$ to the intermediate state $\ket{e}=\ket{6P_{3/2}, F=3}$, and a strong \qty{1013}{nm} coupling laser driving the transition between $\ket{e}$ and the Rydberg state $\ket{r} = \ket{70S_{1/2}}$, with the corresponding energy level diagram shown in Fig.~\ref{Fig2}(b). The probe laser beam has a \qty{1.5}{mm} diameter with \qty{1}{mW} power, while the coupling beam diameter is \qty{2.7}{mm} with tunable power $P_\mathrm{c}$ ranging from 16 to \qty{190}{mW}.

Although there are many Zeeman sublevels, the probe and coupling lasers can simultaneously couple to these energy levels. When both lasers are linearly polarized, each can be decomposed into superpositions of left- ($\sigma^-$) and right-circularly ($\sigma^+$) polarized components. According to the transition selection rule, there exist closed loops in which a combination of $\sigma^+$-$\sigma^-$ or $\sigma^-$-$\sigma^+$ cascaded probe-coupling transition can interfere, as highlighted by the red and blue arrows in Fig.~\ref{Fig2}(b). Then, two linearly polarized lasers can simultaneously form many close loops, and these closed loops acquire the same synthetic gauge phase $\theta$ that is determined by the relative polarization angle between the probe and coupling fields. To implement precise polarization control, we employ polarization beam splitters (PBSs) to purify the initial polarization states of both beams. Half-wave plates (HWP) then allow continuous rotation of the linear polarization orientation, while a quarter-wave plate (QWP) enables switching to circular polarization for control experiments. The two beams counterpropagate through a heated vapor cell and are subsequently separated by a dichroic mirror (DM). The transmitted probe signal ($U$) is detected by a silicon avalanche photodiode (PD).

To study the Rydberg EIT, we first calibrate the probe laser detuning through the saturated absorption spectrum (SAS)~\cite{preston1996dopplerfree}, as shown in the top panel of Fig.~\ref{Fig2}(c). However, the EIT signal cannot be directly detected for the weak transition [bottom panel of Fig.~\ref{Fig2}(c)]. Thus, we employ the lock-in amplification technique~\cite{drever1983laser}, where a \qty{130}{kHz} amplitude modulation is introduced to the coupling laser through an acousto-optic modulator, and the EIT signal-to-noise ratio can be enhanced by extracting the modulation of the transmission. Figure~\ref{Fig2}(d) presents the experimentally observed Rydberg EIT spectrum with $P_\mathrm{c}=$\qty{100}{mW}, where the probe light is horizontally (H) polarized while the coupling light is circularly polarized, avoiding the formation of closed loops of transitions. Here, the EIT spectral profile shows distinct asymmetry, attributed to Rydberg interactions between atoms~\cite{zhangMicrowavecoupledOpticalBistability2025}, as predicted in Fig.~\ref{Fig1}(d).



\begin{figure}
\centering

\includegraphics[scale=1]{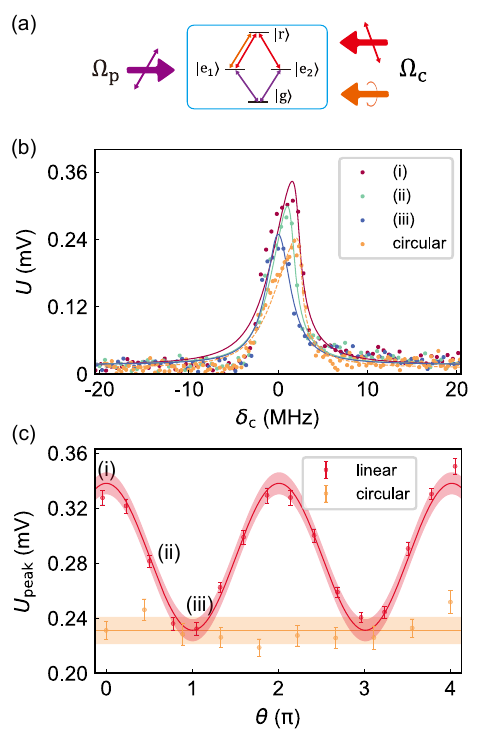}
\caption{(a) Level diagram depicting excitation by a linear probe (purple) and a linear (red) or left circular (orange) coupling field. (b) EIT spectra for (i) parallel (H-H), (ii) intermediate, and (iii) perpendicular (H-V) alignment between the linear polarization directions of coupling and probe lasers, and the circular polarized coupling laser EIT spectra as shown. (c) EIT peak amplitude ($U_\mathrm{peak}$) versus the gauge phase $\theta$, which is derived as two times of the relative polarization angle between coupling and probe lasers.}
\label{Fig3}
\end{figure}

\smallskip
\noindent
\textit{4. Gauge phase.-}
With this experimental setup, we systematically investigated the polarization dependence of the Rydberg EIT to reveal the synthetic gauge phases. Figure~\ref{Fig3}(a) depicts the probe and control laser polarizations for verifying the gauge phase. With the probe laser fixed at H polarization, we systematically rotate the angle of the linearly polarized coupling laser using a motorized half-wave plate and record the EIT spectrum at each angle, in contrast with the results in Fig.~\ref{Fig3}(b), where the coupling is circularly polarized. As shown in Fig.~\ref{Fig3}(b), the curves (i)-(iii) show the results for the angle between the control and probe polarization being 0 (parallel, H-polarized), $\pi/4$, and $\pi/2$ (perpendicular, V-polarized), with a smaller angle showing a weaker EIT peak value.


\begin{figure*}
\centering
\includegraphics[scale=1]{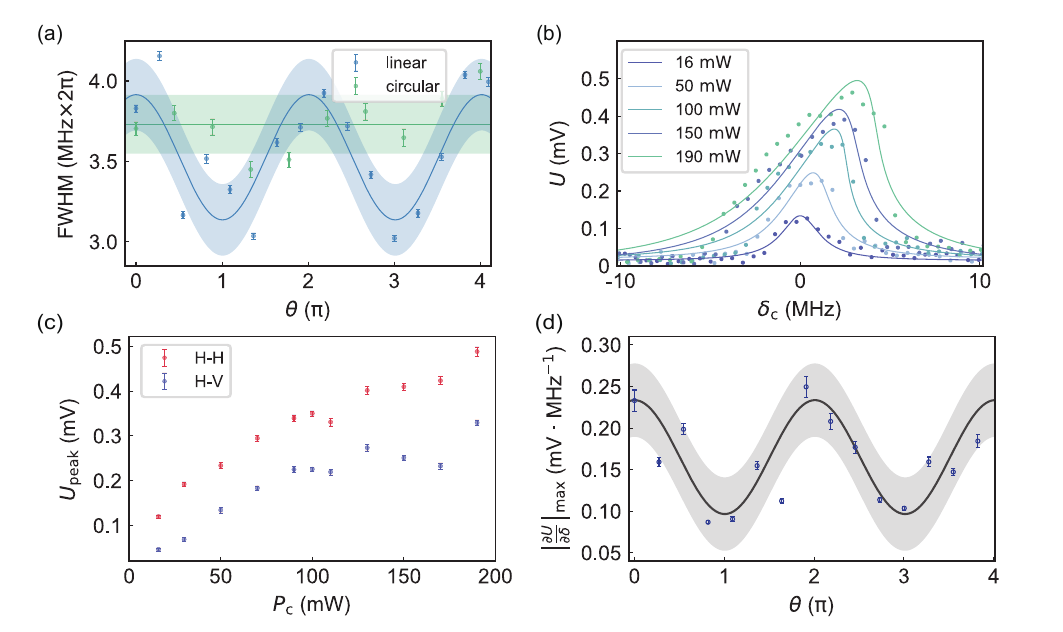}
\caption{(a) EIT linewidth versus $\theta$ for linear (blue) and circular (red) polarized coupling lasers. (b) Rydberg EIT spectra (H-H polarization configuration) for different coupling laser powers $P_\mathrm{c}$. (c) Rydberg EIT peak transmission versus $P_\mathrm{c}$ for parallel (H-H, blue) and perpendicular (H-V, red) polarization configurations. (d) Maximum spectral slop versus $\theta$.}
\label{Fig4}
\end{figure*}

This modulation confirms that the relative polarization angle governs the EIT transmission. According to the transitions shown in Fig.~\ref{Fig1}(a), we derived that the gauge phase $\theta$ equals two times of the angle between the two linearly polarized lasers, irrespective of the direction of the polarization of either probe or coupling laser. Thus, we obtain $\theta=0$ for parallel polarizations, as both transition paths from the ground state to the Rydberg state have the same phase, while $\theta=\pi$ for perpendicular polarizations, and the two transition paths exhibit destructive interference. 

Through carefully scanning the angle, we obtained the dependence of the EIT peak voltage as a function of the gauge phase $\theta$. As shown in Fig.~\ref{Fig3}(c), the experimental data show sinusoidal oscillations and agree with theoretical expectations. To unambiguously verify that the observed oscillations originate from the closed-loop transition geometry, we perform an experiment by switching the input coupling laser from initially linear polarization to right-handed circular polarization, while rotating the direction of the linearly polarized probe laser. Then, the EIT peak value has a negligible dependence on the gauge phase, confirming that the synthetic gauge phase is generated by the closed-loop transitions induced by two linearly polarized lasers. 

The feature of the Rydberg interactions among atoms is investigated in Fig.~\ref{Fig4}. First, for the configuration of Fig.~\ref{Fig3}, the EIT linewidth also shows a sinusoidal oscillation with gauge phase for two linearly polarized inputs, while the oscillation is hardly visible from experimental data for a circularly polarized control laser. The synchronized oscillation of the linewidth and EIT peak confirms that the stronger Rydberg population in the atomic ensemble leads to a broadening of the transition. In Fig.~\ref{Fig4}(b), we examine the interplay between the peak value and the linewidth of EIT lineshape by increasing the $P_\mathrm{c}$ from \qty{16}{mW} to \qty{190}{mW}, where both probe and coupling lasers are H-polarized. As $P_\mathrm{c}$ increases, the peak value becomes more significant and the linewidth broadens. 

The effect of the synthetic gauge phase on the Rydberg interactions is studied in Fig.~\ref{Fig4}(c), by comparing the EIT transmission for $\theta=0$ (parallel polarizations) and $\theta=\pi$ (perpendicular polarizations) across a range of coupling powers. Throughout the entire power range, the $\theta=0$ case consistently yields higher transmission, and the effect is more pronounced in the high-power regime where the Rydberg interaction is stronger. Figure~\ref{Fig4}(d) provides a more quantitative presentation of the Rydberg interaction-induced nonlinear asymmetric spectral response in EIT, i.e., the maximum slope $\left| {\mathrm{\partial}T}/{\mathrm{\partial}\delta} \right|_{\mathrm{max}}$. The results reveal a clear sinusoidal modulation with $\theta$, validating our theoretical prediction in Fig.~\ref{Fig1}(f) and demonstrating that the gauge phase actively regulates the effective Rydberg nonlinearity.

\smallskip
\noindent\textit{5. Conclusion.-} 
We have demonstrated a synthetic gauge phase in the Rydberg transitions of a rubidium atomic ensemble. The gauge phase modulates the Rydberg EIT spectral profile and effectively tunes the Rydberg nonlinearity simply by varying the laser linear polarization direction, without requiring changes to laser intensity or frequency. The experiments with a circularly polarized coupling laser validate that closed-loop transitions are essential for observing such dependence on gauge phase. Therefore, our work reveals polarization as an experimental control knob for manipulating both light-matter interactions and Rydberg atom-atom interactions. This synthetic gauge phase establishes a powerful control mechanism for exploring quantum simulation and topological phenomena~\cite{fontaineObservationBogoliubovDispersion2018,zhangObservationParityTimeSymmetry2016,liuBifurcationTimeCrystals2025,chenContinuousSymmetryBreaking2023,goldman2016topological,bernienProbingManybodyDynamics2017,ebadiQuantumPhasesMatter2021,schollQuantumSimulation2D2021,blochManybodyPhysicsUltracold2008}.

\section*{Acknowledgments}
\begin{acknowledgments}
This work was funded by the National Key R\&D Program (Grant No.~2021YFA1402004), the Natural Science Foundation of Anhui Province (Grant No. 2408085QA017) and the National Natural Science Foundation of China (Grants No.~92465201 and 12504569). Y.-N.L. was also supported by China Postdoctoral Science Foundation (Grant No.~2024M753081). This work was also supported by the Fundamental Research Funds for the Central Universities and USTC Research Funds of the Double First-Class Initiative. The numerical calculations in this paper have been done on the supercomputing system in the Supercomputing Center of USTC. This work was partially carried out at the USTC Center for Micro and Nanoscale Research and Fabrication.
\end{acknowledgments}

\bibliographystyle{Zou}
\bibliography{main}
\end{document}